\documentclass[twocolumn,preprintnumbers,secnumarabic,amsmath,amssymb,nofootinbib,floatfix]{revtex4}
\usepackage{varioref,exscale,latexsym,amsmath,amssymb}
\usepackage{graphicx}

\usepackage{graphicx}
\usepackage{dcolumn}
\usepackage{bm}

\def\beq{\begin{equation}}

\def\eeq{\end{equation}}

\def\beqa{\begin{eqnarray}}

\def\eeqa{\end{eqnarray}}

\begin{document}

\title{{\bf Breaking Diffeomorphism Invariance and Tests for the Emergence of Gravity }}

\medskip\
{\author{Mohamed M. Anber}
\email[Email: ]{manber@physics.umass.edu}
 \author{Ufuk Aydemir}
 \email[Email: ]{uaydemir@physics.umass.edu}
\author{ John F. Donoghue}
\email[Email: ]{donoghue@physics.umass.edu}
\affiliation{Department of Physics,
University of Massachusetts\\
Amherst, MA  01003, USA}

\begin{abstract}
If general relativity is an emergent phenomenon, there may be small violations of diffeomorphism invariance. We propose a phenomenology of perturbatively small  violations of general relativity by the inclusion of terms which break general covariance. These can be tested by matching to the Parameterized Post Newtonian (PPN) formalism. The most sensitive tests involve pulsar timing and provide an extremely strong bound, with a dimensionless constraint of order $10^{-20}$ relative to gravitational strength.
\end{abstract}
\maketitle


\section{Introduction}
Most approaches to new theories invoke higher gauge symmetries beyond those seen at low energies, with the known symmetries being respected at all energies. However, it is also possible that the gauge theories in nature are emergent, low energy manifestations from a fundamental theory with quite different degrees of freedom and perhaps without an intrinsic gauge invariance. Emergent phenomena are common in nature. While there may be obstacles to emergent mechanisms for fundamental theories, the possibility certainly deserves study.

Gravitational physics is a good candidate for an emergent theory because of the poor high energy behavior of general relativity. While the low energy theory forms a good quantum effective field theory\cite{eft} , at high energies the perturbative theory falls apart. This may signal the need for new degrees of freedom and new interactions beyond the Planck scale. In some theories, those with a background independence or explicit general covariance, the emergent gravitational theory will fully respect diffeomorphism invariance\cite{background}. However, other candidate may not have this feature. There are several attempts to produce emergent gauge theories and gravity \cite{emergent}, with many using ideas based on condensed matter emergent analogies, as well as recent work on Horava-Lifshitz theories\cite{Horava}. Indeed, the Witten-Weinberg theorem\cite{wittenweinberg} implies that if emergence is to explain all the gauge theories, spacetime and Lorentz invariance may need to be emergent.

Emergent theories that start from a framework without diffeomorphism invariance will leave behind an imprint of the lack of this invariance. By definition, these theories reduce to general relativity at low energies, plus small corrections. At the very least, loop diagrams probe the very highest energies and will be sensitive to the lack of the invariance at the fundamental scale.

Gravity is also a good place for emergent phenomenology.  The gravitational interactions are weak, suppressed by powers of the Planck scale. Therefore, it is plausible that the residual effects of non-invariance will be relatively more visible in gravitational interactions.

Motivated by these considerations, we will initiate a phenomenological study of potential small breaking of diffeomorphism invariance. While we do not know the underlying theory and therefore do not know the magnitude and form of the symmetry breaking, we study an effective Lagrangian with terms that break diffeomorphism invariance, and proceed to match on to the Parameterized Post-Newtonian (PPN) framework in order to provide a remarkably stringent bound.

The plan of this work is as follows: In Sec. 2, we discuss the infinite set of operators which violate diffeomorphism invariance and describe a set of terms which involve two derivatives of the metric which we will use as our test cases for this study. In Sec 3, we look at the linearized theory, find the propagator and calculate the bending of light. Because gauge non-invariant theories have extra degrees of freedom and could have ghosts we use the propagator analysis to probe for the existence of ghosts. We find that certain values of the parameters are required in order to be ghost-free. In Sec. 4, we start the matching to the PPN framework, which is carried out in detail in Sec. 5-8. In Sec. 9 we describe the resulting phenomenological constraint. Sec 10 is a brief summary. Several appendices describe auxiliary features of our treatment.

\section{Formulation}

Once one opens up the action to include non-covariant terms, the possibilities are myriad. When we organize the theory in an energy expansion, the action is ordered by powers of derivatives. In a theory with a metric, the metric $g_{\mu\nu}(x)$ is the primary field.  There are no Lorentz invariant combinations of the metric without any derivatives, aside from the cosmological constant. In this case, the leading possible non-covariant terms in the action start with two derivatives. Terms with four derivatives or more would be suppressed at low energy. In this paper we study the two derivative modifications to the action.

One should recognize that there might be other ways to test diffeomorphism breaking. For example, one can also include terms which violate Lorentz invariance\cite{kostelecky}, as such breaking is also a form of diffeomorphism violation. Indeed there might be good reasons for including such terms for an emergent theory\cite{gauge paper}. The breaking of Lorentz invariance has been already been studied in gravity \cite{bailey} and we will study covariance breaking terms which are Lorentz invariant.

Moreover, if one gives up covariance and gives a special weight to a flat metric, it is possible that one could  consider $h_{\mu\nu}(x)=g_{\mu\nu}(x)-\eta_{\mu\nu}$ as the primary field. In this case, there could be terms with zero derivatives in the actions with the leading effect being the Pauli-Fierz mass term \cite{Fierz:1939ix}. This term has been heavily studied and is phenomenologically ruled-out at essentially any magnitude. When the mass is bigger than a critical fraction of the curvature, the van Dam-Veltman -Zakharov (vDVZ) discontinuity \cite{vanDam:1970vg, Zakharov:1970cc} says that the predictions drastically disagree with general relativity. When the mass is smaller than the critical curvature the mass generates an intrinsic instability \cite{instability} in the spacetime in flat, de Sitter or Freidman-Robertson-Walker cosmology,. Only anti-de Sitter spaces escape these serious problems. However, anti-de Sitter space does not appear to be selected in Nature. So it appears that this form of diffeomorphism breaking must be identically zero. Combinations with higher powers of $h_{\mu\nu}(x)$ and zero derivatives should also be studied. Some of these are listed in Appendix A. However,  we will turn our attention to the next order in the derivative expansion.

In the derivative expansion the next terms that would occur would be those with two derivatives of the metric. These are the ones that we study below. In addition there could be others with four derivatives of the metric. In principle these would be suppressed at low energy since the derivatives turn into factors of the graviton energy. Since we have no prior knowledge of the mass scale appearing in the energy expansion, these should be studied as well, but we reserve this for future work. We provide a classification of the symmetry breaking operators in the linear approximation up to sixth order in Appendix A.

Given these possibilities, we do not attempt a fully general analysis, but will look at some possibilities which have not been studied before and for which we can obtain a particularly tight bound. More general possibilities will be considered in future work.

In this section, we introduce a general second derivative Lagrangian involving the connection in ways that break the diffeomorphism invariance.  This is a purely dynamical metric theory of gravity which assumes that \cite{Thorne} that there exists a symmetric metric. In addition, we also assume that all non-gravitational fields couple universally to the gravitational field.

Our general action takes the form
\begin{equation}\label{mail Lagrangian}
S=\int d^4 x\,\sqrt{-g}\,{\cal L}\,.
\end{equation}
where
\begin{equation}
{\cal L}=\frac{1}{16\pi G}\big[R+\sum_{i=1}^{7}a_i{\cal L}_i\big]+{\cal L}_m\,.
\end{equation}
The first term above is the usual Einstein-Hilbert term, ${\cal L}_m$ is the matter Lagrangian, while ${\cal L}_i$ are the diffeomorphism-violating pieces
\begin{eqnarray}\label{operatorbasis}
\nonumber
{\cal L}_1&=&-\,g^{\mu\nu}\Gamma^\alpha_{\mu\lambda}\Gamma^{\lambda}_{\nu\alpha}\,,\quad {\cal L}_2=-\,g^{\mu\nu}\Gamma^\alpha_{\mu\nu}\Gamma^{\lambda}_{\lambda\alpha}\\
\nonumber
{\cal L}_3&=&-\,g^{\alpha\gamma}g^{\beta\rho}g_{\mu\nu}\Gamma^\mu_{\alpha\beta}\Gamma^\nu_{\gamma\rho}\,,\quad {\cal L}_4=-\,g^{\alpha\gamma}g_{\beta\lambda}g^{\mu\nu}\Gamma^\lambda_{\mu\nu}\Gamma^\beta_{\gamma\alpha}\\
\nonumber
{\cal L}_5&=&-g^{\alpha\beta}\Gamma^{\lambda}_{\lambda\alpha}\Gamma^{\mu}_{\mu\beta}\,,\quad
{\cal L}_6=-g^{\mu\nu}\partial_{\nu}\Gamma_{\mu\lambda}^{\lambda}\\
{\cal L}_7&=&-g^{\mu\nu}\partial_{\lambda}\Gamma_{\mu\nu}^{\lambda}\,,
\end{eqnarray}
and $\{a_i\}$ are small coefficients. The Greek indices run over four spacetime coordinates, and we take the Lorentz signature to be mostly positive. Notice that ${\cal L}_{EH}$ can be written as $-{\cal L}_1+{\cal L}_2+{\cal L}_6-{\cal L}_7$. Using integration by parts and the identities
\begin{equation}\label{important identities} \partial_{\nu}g^{\rho\gamma}=-g^{\rho\alpha}\Gamma^{\gamma}_{\nu\alpha}-g^{\beta\gamma}\Gamma^{\rho}_{\beta\nu}\,,\partial_{\mu}\sqrt{-g}=\sqrt{-g}\Gamma_{\mu\nu}^{\nu}
\end{equation}
we obtain
\begin{eqnarray}
{\cal L}_6={\cal L}_2+\mbox{\scriptsize surface term}\,, \quad {\cal L}_7=2{\cal L}_1-{\cal L}_2+\mbox{\scriptsize surface term}\,.
\end{eqnarray}
Hence, ${\cal L}_6$ and ${\cal L}_7$ are not independent and we drop them in our analysis. Moreover, using the same identities, one can show that any other term of square derivative, e.g. $\sqrt{-g}\partial_{\gamma}g^{\alpha\beta}\partial^{\gamma}g_{\alpha\beta}$ or $\sqrt{-g}\partial_{\alpha}\partial_{\beta}g^{\alpha\beta}$, can be uniquely expressed as a linear combination of ${\cal L}_{1}$ through ${\cal L}_{5}$ modulo surface terms. This set is not unique beyond linear order, as once one has given up covariance one can have an infinite set of Lagrangians of the form $(\sqrt{-g})^n{\cal L}$ by adding extra powers of $(\sqrt{-g})^n$ starting at the same two derivative order. However, such modifications do not contribute new terms to the linear equations of motion.

The use of the connection in Eq. \ref{operatorbasis} is important for our matching to the PPN formalism. As we will see in the next section, this basis is larger than needed for the linear analysis keeping only the leading term in the expansion of the metric. However, the PPN formalism is sensitive to the higher order terms in the expansion of the metric. By using the connection in our operator basis, these higher order terms are part of expansion of the connection.

\section{Linearized version of the equations of motion}

Writing $g_{\mu\nu}=\eta_{\mu\nu}+h_{\mu\nu}$, and retaining only the quadratic contribution of $h$, we obtain
\begin{eqnarray}
\nonumber
{\cal L}_1^{(2)}&=&-\frac{1}{4}\left(-T_1+2T_2 \right)\,,\quad {\cal L}_2^{(2)}=-\frac{1}{4}\left(2T_3-T_4 \right)\\
\nonumber
{\cal L}_3^{(2)}&=&-\frac{1}{4}\left(3T_1-2T_2 \right)\,,\quad {\cal L}_4^{(2)}=-\frac{1}{4}\left(4T_2-4T_3+T_4 \right)\\
\label{expression of lagrangians}
{\cal L}_5^{(2)}&=&-\frac{T_4}{4}\,,
\end{eqnarray}
where
\begin{eqnarray}
\nonumber
T_1&=&\partial_{\gamma}h_{\alpha\beta}\partial^{\gamma}h^{\alpha\beta}\,,\quad T_2=\partial_{\gamma}h_{\alpha\beta}\partial^{\beta}h^{\alpha\gamma}\\
\nonumber
T_3&=&\partial_{\alpha}h\partial_{\beta}h^{\alpha\beta}\,,\quad T_4=\partial_{\alpha}h\partial^{\alpha}h\,.
\end{eqnarray}

Inspection of eq. (\ref{expression of lagrangians}) reveals that the quadratic Lagrangians ${\cal L}_1^{(2)}$ to ${\cal L}_4^{(2)}$ are independent. Hence, we conclude that the Lagrangians ${\cal L}_{1}$ to ${\cal L}_{4}$ are also independent.

Variation of the quadratic version of the Lagrangian (\ref{mail Lagrangian}) results in the linearized equations of motion
\begin{eqnarray}
\nonumber
&&\left(-1-a_1+3a_3\right)\square h^{\alpha\beta}+\left(1+a_1-a_3+2a_4 \right)\left(\partial^\alpha\partial_\gamma h^{\beta\gamma}\right.\\
\nonumber
&&\left.+\partial^\beta\partial_\gamma h^{\alpha\gamma} \right)+\left(-1+a_2-2a_4  \right)\eta^{\alpha\beta}\partial_\mu\partial_\nu h^{\mu\nu}\\
\nonumber
&&+\left(-1+a_2-2a_4 \right)\partial^{\alpha}\partial^{\beta}h+\left(1-a_2+a_4+a_5 \right)\eta^{\alpha\beta}\square h\\
\label{com linearized equation of motion}
&&=16\pi G T^{\alpha\beta}\,,
\end{eqnarray}
where $h=h^{\alpha}_{\alpha}$, and $T^{\alpha\beta}$ is the energy momentum tensor that results by varying the matter action with respect to $g_{\alpha\beta}$.

Throughout this work we assume that the energy momentum tensor of the matter fields is conserved. This translates in the linear theory into $\partial^{\alpha}T_{\alpha\beta}=0$. Taking the derivative of eq. (\ref{com linearized equation of motion}) we obtain
\begin{eqnarray}
\nonumber
&&2\left(a_3+a_4\right)\square \partial_{\beta}h^{\alpha\beta}+\left(a_1+a_2-a_3 \right)\partial^{\alpha}\partial_{\mu}\partial_{\nu}h^{\mu\nu}\\
\label{general constraint equation}
&&+\left(a_5-a_4 \right)\partial^\alpha\square h=0\,.
\end{eqnarray}
This equation is satisfied provided that we have either

\begin{itemize}

 \item {\em (i)} Trivial case: all the coefficients $a_1$ to $a_5$ are set equal to zero, or they satisfy $a_4=a_5$, $a_3+a_4=0$ and $a_1+a_2-a_3=0$ ( the last two conditions result in $a_1+a_2-3a_3-2a_4=0$, see the comments below eq. (\ref{projection of propagator})). This is the  case of a diffeomorphism-invariant theory (in this case general relativity GR).

 \item {\em (ii)} General case: all coefficients $a_1$ to $a_5$ are different from zero. In this case the realization of eq. (\ref{general constraint equation}) can be guaranteed  if we impose the constraints
\begin{equation}\label{linearized constraint}
\partial_{\alpha}h^{\alpha\beta}=0\,, \quad \mbox{and}\quad  a_4=a_5.
\end{equation}
The physical consequences to linear order (light bending) of this case is worked out below.

 \item {\em (iii)} Special case I: all coefficients are set to zero except $a_4 \neq0$. In this case  we make use of the field redefinition $h^{\alpha\beta}=\bar h^{\alpha\beta}-\eta^{\alpha\beta}\bar h/2$ to bring eq. (\ref{general constraint equation}) to the form
\begin{eqnarray}
\nonumber
&&-\square \bar h^{\alpha\beta}+(1+2a_4)\left(\partial^{\alpha}\partial_{\gamma}\bar h^{\beta\gamma}+\partial^{\beta}\partial_{\gamma}\bar h^{\alpha\gamma} \right)\\
\label{linearized for a4}
&&-(1+2a_4)\eta^{\alpha\beta}\partial_{\mu}\partial_{\nu}\bar h^{\mu\nu}=16\pi G T^{\alpha\beta}\,.
\end{eqnarray}
Then, imposing the condition $\partial_{\alpha}T^{\alpha\beta}=0$ we get the constraint $\partial_{\alpha}\bar h^{\alpha\beta}=0$. Using this constraint in eq. (\ref{linearized for a4}) we obtain $\square \bar h^{\alpha\beta}=16\pi G T^{\alpha\beta}$. Hence, to linear order, setting $a_4\neq 0$ does not lead to any physical consequences beyond GR.

 \item {\em (iv)} Special case II: all the coefficients are set to zero except $a_5 \neq0$. In this case one can use the transformation $h^{\alpha\beta}=\bar h^{\alpha\beta}-\eta^{\alpha\beta}\bar h/4$ to eliminate $a_5$. Again, this leads to no physical consequences on the linear level beyond GR.
\end{itemize}

It is important to notice here what has been stated in the literature that if we want a more general equation than Einstein's equation that respects Lorentz symmetry and reduces in the weak field limit to a second order equation, then we have to include other elements that are unrelated to the metric tensor or its derivatives, and we must give up the possibility of deriving Newton's theory as a limiting case (see e.g. \cite{weinberg gravitation}). The above statement is true {\em only} if we do not give up the diffeomorphism invariance as a fundamental symmetry of the underlying  manifold. However, we have shown that breaking this symmetry can still result in a second order equation that has Newton's theory as a limiting case (see the discussion below) provided that we take the covariance breaking coefficients to be small enough.

Also, one may argue that we should run into troubles once we break the diffeomorphism invariance. The classical example is Pauli-Fierz massive gravity \cite{Fierz:1939ix}. The argument is that since diffeomorphism symmetry is a dynamical symmetry, breaking it will excite the scalar modes that become strongly coupled even when we send the graviton mass to zero. This is the famous van Dam-Veltman-Zakharov (vDVZ) discontinuity \cite{vanDam:1970vg,Zakharov:1970cc}. As we show below, this kind of discontinuity does not happen in our case. A simple explanation can be given here (see e.g. the first reference in \cite{instability}).  The equation of motion for Pauli-Fierz massive gravity reads
\begin{eqnarray}
\nonumber
&&-\square h^{\alpha\beta}+\left(\partial^\alpha\partial_\gamma h^{\beta\gamma}
+\partial^\beta\partial_\gamma h^{\alpha\gamma} \right)-\eta^{\alpha\beta}\partial_\mu\partial_\nu h^{\mu\nu}\\
\nonumber
&&-\partial^{\alpha}\partial^{\beta}h+\eta^{\alpha\beta}\square h+m^2\left(h^{\alpha\beta}-\eta^{\alpha\beta}h \right)\\
\label{pauli feirz}
&&=16\pi G T^{\alpha\beta}\,.
\end{eqnarray}
Taking the divergence and trace of the above equation results in the five constraints
\begin{equation}\label{pauli feirz constraint}
\partial_\alpha h^{\alpha\beta}=\partial^\beta h\,, h=\frac{16\pi G}{3 m^2}T\,.
\end{equation}
Along with the obvious behavior of $1/m^2$ in the limit of zero mass, the first constraint in (\ref{pauli feirz constraint}) does not reduce to a guage condition consistent with equations of motion in the limit $m=0$. This behavior, which signals the presence of a problem, is absent in our case since the constraint $\partial_\alpha h^{\alpha\beta}=0$ reduces to a gauge condition as we send $a_1$, $a_2$, etc. to zero.

\subsection{The propagator and bending of light}

To further study the linearized theory, it is instructive to write down the graviton propagator. This can be accomplished by writing the quadratic Lagrangian in the form $h_{\mu\nu}O^{\mu\nu,\alpha\beta}h_{\alpha\beta}$, and finding the inverse of the operator $O$. To impose the constraint $\partial^{\alpha}h_{\alpha\beta}=0$, we insert in the Lagrangian  a term $\Lambda\left(\partial^{\alpha}h_{\alpha\beta} \right)^2$, and take the limit $\Lambda\rightarrow \infty$ at the end of calculations. Hence, we find
\begin{eqnarray}
\nonumber
D_{\mu\nu,\rho\sigma}(k)&=&-A\eta_{\mu\nu}\eta_{\rho\sigma}/k^2+B\left(\eta_{\mu\rho}\eta_{\nu\sigma}+ \eta_{\mu\sigma}\eta_{\nu\rho} \right)/k^2\\
\nonumber
&&+A\left(\eta_{\mu\nu}k_\rho k_\sigma+\eta_{\rho\sigma}k_\mu k_\nu \right)/k^4\\
\nonumber
&&-B\left(\eta_{\mu\rho}k_\nu k_\sigma+\eta_{\mu\sigma}k_\nu k_\rho+\eta_{\nu\sigma}k_\mu k_\rho\right.\\
\label{propagator for diff gravity}
&&\left.+\eta_{\nu\rho}k_\mu k_\sigma \right)/k^4+Ck_\mu k_\nu k_\rho k_\sigma/k^6\,,
\end{eqnarray}
where the constants $A$, $B$, $C$, and $D$ are given by
\begin{eqnarray}
\nonumber
A&=&\frac{1-a_2+2a_4}{\left(1+a_1-3a_3  \right)\left(2-a_1-3a_2+3a_3+6a_4 \right)}\\
\nonumber
B&=&\frac{1}{2\left(1+a_1-3a_3 \right)}\\
C&=&\frac{1-a_1-2a_2+3a_3+4a_4}{\left(1+a_1-3a_3  \right)\left(2-a_1-3a_2+3a_3+6a_4 \right)}
\end{eqnarray}
and we have put $a_4=a_5$. As expected, the propagator is continuous to the GR result as we set $a_1=a_2=...a_4=0$.

Now consider two particles with conserved energy momentum tensor $T^{\mu\nu}_{(1)}$ and $T^{\rho\sigma}_{(2)}$ interacting via the exchange of a graviton. The scattering amplitude is given by
\begin{eqnarray}
\nonumber
&&GT^{\mu\nu}_{(1)}(k)D_{\mu\nu,\rho\sigma}(k)T^{\rho\sigma}_{(2)}(k)\\
&&=\frac{G}{k^2}\left(-A T_{(1)}T_{(2)}+2BT^{\mu\nu}_{(1)}T_{(2)\,\mu\nu} \right)\,.
\end{eqnarray}
The scattering amplitude between to chunks of non-relativistic matter is proportional to
\begin{equation}
\frac{G_{\mbox{\scriptsize eff}}}{2k^2}T^{00}_{(1)}T^{00}_{(2)}\,
\end{equation}
where $G_{\mbox{\scriptsize eff}}=2G(2B-A)$. Now, taking $T^{\mu\nu}_{(1)}$ and $T^{\mu\nu}_{(2)}$ to be respectively the energy momentum tensor of the sun and photon we obtain the scattering amplitude, and hence the light bending effect
\begin{equation}\label{propagator light bending}
\frac{G_{\mbox{\scriptsize eff}}}{2k^2}\frac{2-a_1-3a_2+3a_3+6a_4}{1-a_1-2a_2+3a_3+4a_4}\,.
\end{equation}
In appendix D, we obtain the same results using the PPN formalism.

\subsection{Ghost Analysis}

Finally, we consider the issue of a possible ghost instability in diffeomorphism-violating theories of gravity. Generally, ghosts can appear in effective field theories. However, they usually show up only at energies above the cut-off scale of the theory. A thorough analysis of the linearized version of the action (\ref{mail Lagrangian}) to search for such instabilities was given in \cite{Alvarez:2006uu} in the context of transverse-diffeomorphism theories of gravity \cite{Alvarez:2005iy}. Still, one can learn about these instabilities by studying the behavior of the momentum space propagator. A ghost propagator will have the opposite sign of a healthy degree of freedom. Hence, existence of an instability can be read from (\ref{propagator for diff gravity}) as terms with the wrong sign. A quick way of doing this is by sandwiching the propagator between the energy momentum tensors of two sources as we did above, and then projecting out the transverse traceless part of the spin-2 particle. Hence, the scattering amplitude can be written as
\begin{eqnarray}
\nonumber
&&\frac{2GB}{k^2}\left[T^{\mu\nu}_{(1)}T_{(2)\,\mu\nu}-\frac{1}{2}T_{(1)}T_{(2)} \right]\\
\label{projection of propagator}
&&-\frac{GB}{k^2}\frac{a_1+a_2-3a_3-2a_4}{2-a_1-3a_2+3a_3+6a_4}T_{(1)}T_{(2)}\,.
\end{eqnarray}
The first term is the usual spin-2 graviton coupled to matter, while the second term represents massless interaction between
conserved sources. This massless degree of freedom has a healthy kinetic term (not a ghost) provided that $a_1+a_2-3a_3-2a_4\leq0$. The saturation of this inequality decouples the massless mode.

On the other hand, the analysis drawn in \cite{Alvarez:2006uu} showed that the Minkowskian vacuum in the linear version of (\ref{mail Lagrangian}) admits a linear classical instability for the vector modes unless we impose the constraint $a_3+a_4=0$. These modes couple to the derivative of the energy momentum tensor, and hence they do not show up in (\ref{projection of propagator}) since we are considering conserved sources.The constraint $a_3+a_4=0$ along with the saturation of the above inequality ($a_1+a_2-3a_3-2a_4=0$)
restores the  diffeomorphism invariance of the theory (GR)

We show below that experimental tests of gravity bound the values of $\{a_i\}$ to be of order $10^{-20}$. This suppresses any significance for ghosts or other classical instability, if any, at low energies. At higher energies, one expects higher order terms, possibly higher derivative Lagrangians, to contribute new degrees of freedom rendering the UV theory ghost free.

\section{Nonlinear equations of motion}

The linear approximation is not sufficient for the PPN analysis that we are about to undertake. The iteration that is involved in that formalism mixes different powers of the linear field $h_{\mu\nu}$. Because our original Lagrangian was written in terms of the connection, we are able to include higher order terms using the original operators. Moreover at this stage we are going to restrict our treatment to one of the operators in our basis, ${\cal L}_3$. While we plan to report on it in future work, the treatment of the general case is very cumbersome and the use of this operator is sufficient to identify the strongest test, coming from the preferred frame parameter $\alpha_3$, and obtain a very strong constraint.

The nonlinear equations of motion of the system can be found using the Euler-Lagrange equations
\begin{equation}
\frac{\partial \sqrt{-g}\,{\cal L}}{\partial g_{\alpha\beta}}-\partial_{\mu}\left(\frac{\partial \sqrt{-g}\,{\cal L}}{\partial g_{\alpha\beta,u}}\right)=0\,,
\end{equation}
where the comma denotes ordinary derivative.

 Taking into account the identities  $g_{\alpha\beta}g^{\beta\gamma}=\delta_\alpha^\gamma$, and $\partial\sqrt{-g}/\partial g_{\alpha\beta}=\sqrt{-g}g^{\alpha\beta}/2$ we obtain (from here on, we drop the subscript in $a_3$ to reduce notational clutter)
\begin{equation}\label{main equation}
R_{\mu\nu}-\frac{1}{2}g_{\mu\nu}R+a{\cal M}_{\mu\nu}=8\pi G\, T_{\mu\nu}\,,
\end{equation}
where ${M}_{\mu\nu}={\cal B}_{\mu\nu}+{\cal D}_{\mu\nu}$ , and the functions ${\cal B}_{\mu\nu}$ and ${\cal D}_{\mu\nu}$ are given by
%
\begin{eqnarray}
\nonumber
{\cal B}_{\mu\nu}=&-&\frac{1}{2}g_{\mu\nu}g_{\alpha\beta}g^{\gamma\delta}g^{\epsilon\eta}\Gamma^{\alpha}_{\gamma\epsilon}\Gamma^\beta_{\delta\eta}+g^{\alpha\beta}g^{\gamma\delta}g_{\nu\phi}g_{\mu\epsilon}\Gamma^{\epsilon}_{\alpha\gamma}\Gamma^{\phi}_{\beta\delta}\\
&+&2g^{\phi\epsilon}g^{\alpha\gamma}g_{\delta\epsilon}g_{\phi\beta}\Gamma_{\mu\alpha}^{\beta}\Gamma^{\delta}_{\nu\gamma}\,,
\end{eqnarray}
%
\begin{equation}
{\cal D}_{\mu\nu}=\Gamma^{\lambda}_{\alpha\lambda}{\cal A}^{\alpha}_{\mu\nu}+ {\cal A}^{\alpha}_{\mu\nu,\alpha}\,,
\end{equation}
and
\begin{equation}
{\cal A}_{\mu\nu}^{\alpha}=g^{\alpha\beta}g_{\gamma\mu}\Gamma^{\gamma}_{\nu\beta}+g^{\alpha\beta}g_{\gamma\nu}\Gamma^{\gamma}_{\mu\beta}-\Gamma_{\mu\nu}^{\alpha}\,.
\end{equation}
Since the equation of motion (\ref{main equation}) is not invariant under general coordinate transformations, the existence of a solution requires that we impose a consistency condition, as we did in the linearized case. As usual, we assume that the energy momentum tensor is conserved, i.e. it satisfies $\nabla^{\mu}T_{\mu\nu}=0$, where $\nabla$ denotes the covariant derivative. In addition, we have from the geometry $\nabla^{\mu}G_{\mu\nu}=0$. Hence we must also have
\begin{equation}\label{cons condition}
\nabla^{\mu}{\cal M}_{\mu\nu}=0\,,
\end{equation}
which is the consistency condition.

In the rest of the paper, we use the parametrized post Newtonian (PPN) formalism to bound the numerical value of $a$ by comparing the outcomes of our theory of gravity with the experimental data.

\section{the PPN formalism}

To compare the different theories of gravity to experiments, we take the slow motion and weak field approximation. Such treatment is perturbative in nature and known as the post Newtonian formalism \cite{Will:1993ns}. In this method, one expands in a small expansion parameter which is taken to be the velocity $v$ of a fluid element
\begin{equation}
U\sim v^2\sim p/\rho\sim\Pi\sim {\cal O}(2)\,,
\end{equation}
where $U$ is the Newtonian potential, $p$ is the pressure of the fluid, $\rho$ is its rest mass density, and $\Pi$ is the specific energy density (ratio of energy density to rest-mass density). The power of velocity $v$ is ${\cal O}(1)$, $U$ is ${\cal O}(2)$, $\Pi$ is ${\cal O}(2)$ and $p$ is ${\cal O}(4)$. In addition, since the time evolution of a system is governed by the motion of its constituents, one has $\partial/\partial t \sim \vec v\cdot \nabla$, and hence
\begin{equation}
\frac{|\partial/\partial t|}{|\partial/\partial x|}\sim {\cal O}(1)\,.
\end{equation}

To obtain the Newtonian limit, we need to solve for $g_{00}$ to ${\cal O}(2)$. Assuming $g_{00}\rightarrow 0$ far from the system we find $g_{00}=2 G_{\mbox{\scriptsize eff}} U$. Since we work in units in which the measured gravitational constant is unity, we set
\begin{equation}
G_{\mbox{\scriptsize eff}}\equiv 1\,.
\end{equation}
The post Newtonian corrections to the propagation of light may be found by solving for $ g_{ij}$ to ${\cal O}(2)$. To this order, one can use the linearized equations of motion  (\ref{com linearized equation of motion}). However, for more involved experiments like the perihelion shift of Mercury, we need to know $g_{00}$ to ${\cal O}(4)$. To this order we work out the full PPN parameters using only ${\cal L}_3$ as an example.

In the following, we assume that the matter content is idealized as a perfect fluid, and hence the components of the energy momentum tensor to the relevant order are

\begin{eqnarray}
\nonumber
T^{00}&=&\rho\left(1+\Pi+v^2+2U \right)\\
\nonumber
T^{0i}&=&\rho v^i\\
\nonumber
T^{ij}&=&\rho v^iv^j+p\delta_{ij}\,.\\
\end{eqnarray}
In addition, the metric will be constructed out of few gravitational potentials $U$, $U_{ij}$, $V_i$, $        W_i$, $\Phi_W$, $\Phi_1$, $\Phi_2$, $\Phi_3$, $\Phi_4$, $\Phi_5$, ${\cal A}$, and ${\cal B}$. The reader can refer to appendix B for the explicit form of these potentials as well as the important differential relations they satisfy.

In the next section we review the setup used to solve systematically for $g_{\mu\nu}$ up to ${\cal O}(4)$.

\section{Setup}

Far from the system under investigation, we expect that the metric tensor reduces to that of Minkowski space. Therefore, we expand the metric $g_{\mu\nu}$ about the Minkowskian background $\eta_{\mu\nu}=\{-1,1,1,1\}$ in powers of $v^2$
\begin{eqnarray}
\nonumber
g_{00}&=&-1+\stackrel{(2)}{g_{00}}+\stackrel{(4)}{g_{00}}+...\,,\quad g_{ij}=\delta_{ij}+\stackrel{(2)}{g_{ij}}+\stackrel{(2)}{g_{ij}}+...\\
g_{0i}&=&\stackrel{(3)}{g_{0i}}+\stackrel{(5)}{g_{0i}}+...\,,
\end{eqnarray}
where the Latin indices run over the spatial dimensions, and $g_{\mu\nu}^{(N)}$ is of order $v^N$. If we define the inverse metric as
\begin{eqnarray}
\nonumber
g^{00}&=&-1+\stackrel{(2)}{g^{00}}+\stackrel{(4)}{g^{00}}+...\,,\quad g^{ij}=\delta_{ij}+\stackrel{(2)}{g^{ij}}+\stackrel{(4)}{g^{ij}}+...\\
g^{0i}&=&\stackrel{(3)}{g^{0i}}+\stackrel{(5)}{g^{0i}}+...\,,
\end{eqnarray}
then using the identity $g_{\alpha\beta}g^{\beta\gamma}=\delta_\alpha^\gamma$ we find
\begin{equation}
\stackrel{(2)}{g^{00}}=-\stackrel{(2)}{g_{00}}\,,\quad \stackrel{(2)}{g^{ij}}=-\stackrel{(2)}{g_{ij}}\,,\quad \stackrel{(3)}{g^{0i}}=\stackrel{(3)}{g_{0i}}\,.
\end{equation}
The components of the affine connections are given in appendix C by (\ref{affine connections}). Also, the components of $R_{\mu\nu}$ and ${\cal M}_{\mu\nu}$ have the form
\begin{eqnarray}
\nonumber
R_{00}&=&\stackrel{(2)}{R_{00}}+\stackrel{(4)}{R_{00}}+...\,, \quad {\cal M}_{00}=\stackrel{(2)}{{\cal M}_{00}}+\stackrel{(4)}{{\cal M}_{00}}+...\\
\nonumber
R_{0i}&=&\stackrel{(3)}{R_{0i}}+\stackrel{(5)}{R_{0i}}+...\,, \quad {\cal M}_{0i}=\stackrel{(3)}{{\cal M}_{0i}}+\stackrel{(5)}{{\cal M}_{0i}}+...\\
\nonumber
R_{ij}&=&\stackrel{(2)}{R_{00}}+\stackrel{(4)}{R_{ij}}+...\,, \quad {\cal M}_{ij}=\stackrel{(2)}{{\cal M}_{ij}}+\stackrel{(4)}{{\cal M}_{ij}}+...\,.\\
\label{expansions of M and R}
\end{eqnarray}

At this stage, it is more convenient to take the trace of (\ref{main equation}) and rewrite it in the form
\begin{equation}\label{conv form}
R_{\mu\nu}+a\left({\cal M}_{\mu\nu}-\frac{1}{2}g_{\mu\nu}{\cal M} \right)=8\pi\, G \left(T_{\mu\nu}-\frac{1}{2}g_{\mu\nu}T \right)\,.
\end{equation}
Plugging the expansion (\ref{expansions of M and R}) into eq. (\ref{conv form}) we obtain, to the relevant order,
\begin{eqnarray}
\nonumber
&&\stackrel{(2)}{R_{00}}+\frac{a}{2}\left(\stackrel{(2)}{{\cal M}_{00}}+\stackrel{(2)}{{\cal M}_{ii}} \right)=4\pi G \stackrel{(2)}{T^{00}}\\
\nonumber
&&\stackrel{(2)}{R_{ij}}+a\left( \stackrel{(2)}{{\cal M}_{ij}}+\frac{1}{2}\delta_{ij}\stackrel{(2)}{{\cal M}_{00}}-\frac{1}{2}\delta_{ij}\stackrel{(2)}{{\cal M}_{kk}} \right)=4\pi G \delta_{ij}\stackrel{(2)}{T^{00}}\\
\nonumber
&& \stackrel{(3)}{R_{0i}}+a\stackrel{(3)}{{\cal M}_{0i}}=-8\pi G \stackrel{(3)}{T^{0i}}\\
\nonumber
&& \stackrel{(4)}{R_{00}}+\frac{a}{2}\left(\stackrel{(4)}{{\cal M}_{00}}+\stackrel{(4)}{{\cal M}_{ii}}-\stackrel{(2)}{g_{00}}\stackrel{(2)}{{\cal M}_{ii}}-\stackrel{(2)}{g_{ij}}\stackrel{(2)}{{\cal M}_{ij}} \right)\\
\label{system of expanded equations}
&&= 4\pi G \left(\stackrel{(4)}{T^{00}}+\stackrel{(4)}{T^{ii}}-2\stackrel{(2)}{g_{00}}\stackrel{(2)}{T^{00}} \right)\,,
\end{eqnarray}
where the summation is indicated in ${\cal M}_{ii}$.

Finally, eqs. (\ref{system of expanded equations}) have to be supplemented with the constraint (\ref{cons condition}) to every order, i.e.
\begin{eqnarray}
\nonumber
&&\nabla^i \stackrel{(2)}{{\cal  M}_{ij}}=0\\
\nonumber
&&\nabla^0 \stackrel{(2)}{{\cal M}_{00}}+\nabla^i \stackrel{(3)}{{\cal  M}_{0i}}=0\\
\label{expanded constraint}
&&\nabla^0 \stackrel{(3)}{{\cal M}_{0i}}+\nabla^i \stackrel{(4)}{{\cal  M}_{ij}}=0\,.
\end{eqnarray}
%

\section{Lower order solutions }

In this section, we investigate the Newtonian, $g_{00}^{(2)}$, and the first post Newtonian, $g_{ij}^{(2)}$ and  $g_{0i}^{(3)}$,  limits for our theory of gravity. By direct calculations, we obtain the lower order expressions for $R$ and ${\cal M}$ as given by (\ref{linearized set}) in appendix C. In  addition, the first two constraints in (\ref{expanded constraint}) results in
\begin{equation}\label{constraint}
\stackrel{(2)}{g_{ij,j}}=0\,, \quad \stackrel{(3)}{g_{i0,i}}-\stackrel{(2)}{g_{00,0}}=0\,,
\end{equation}
which greatly simplifies the subsequent analysis.

Plugging eq. (\ref{linearized set}) into eq. (\ref{conv form}) and using  (\ref{constraint}) we find
\begin{eqnarray}
\nonumber
&&(1-3a)\nabla^2 \stackrel{(2)}{g_{00}}=-16\pi G \frac{1+3a}{2+3a}\,\stackrel{(2)}{T_{00}}\\
\nonumber
&&-\stackrel{(2)}{g_{00,ij}}+\stackrel{(2)}{g_{kk,ij}}+(1-3a)\stackrel{(2)}{g_{00,kk}}=16\pi G \frac{\delta_{ij}}{2+3a}\stackrel{(2)}{T}\\
&&-\stackrel{(2)}{g_{00,0i}}+\stackrel{(2)}{g_{kk,0i}}+(1-3a)\stackrel{(3)}{g_{0i,kk}}=-16\pi G \stackrel{(3)}{T_{0i}}\,.
\end{eqnarray}
The  solution of $g_{00}^{(2)}$ is given by
\begin{eqnarray}\label{Lower order solutions}
\stackrel{(2)}{g_{00}}&=&\frac{2G(1+3a)}{(1-3a)(1+3a/2)}U\,.
\end{eqnarray}
Because the PPN formalism works in units in which the gravitational constant is unity, we must set
\begin{equation}
\frac{G(1+3a)}{(1-3a)(1+3a/2)}=1\,.
\end{equation}
Hence, the normalized solutions  of $g_{00}^{(2)}$, $g_{ij}^{(2)}$, and $g_{0i}^{(2)}$  takes the form
\begin{eqnarray}
\nonumber
\stackrel{(2)}{g_{00}}&=&2U\\
\nonumber
\stackrel{(2)}{g_{ij}}&=&\frac{1}{1+3a}\left(U\delta_{ij}+U_{ij} \right)\\
\label{Normalized Lower order solutions}
\stackrel{(3)}{g_{0i}}&=&-\frac{1}{1+3a}\left[(3+6a)V_i+W_i \right]\,.
\end{eqnarray}
%

\section{Higher order solutions}

Finding $g_{00} $ to ${\cal O}(4)$ is a cumbersome step since it involves all the lower order solutions . Moreover, the requirement that the solution satisfies the constraint (\ref{cons condition}) makes it a long and tedious procedure.

The solution of $g_{00}^{(4)}$ can be obtained from the last eq. in (\ref{system of expanded equations}). This involves the higher order perturbations of $R_{00}$, ${\cal M}_{00}$ and ${\cal M}_{ii}$. The expressions for these functions are given in eq. (\ref{higher order terms}). Although we do not need the explicit value of $g_{ij}^{(4)}$ in the PPN formalism, the appearance of this term  in ${\cal M}_{ii}^{(4)}$ necessitates, in general, a simultaneous solution for $g_{ij}^{(4)}$ and $g_{00}^{(4)}$.
\footnote{In fact, it is clear from eq. (\ref{higher order terms}) that we only need the combination $\left( 3\stackrel{(4)}{g_{ii,kk}}-2\stackrel{(4)}{g_{ik,ik}}\right)$.}
 In turn, this adds more to the complexity of the problem by forcing us to feed the system in (\ref{system of expanded equations}) with the $i-j$ equation to ${\cal O}(4)$.  However, since the parameter $a$ is small, we will be interested only in solutions to first order in $a$. To this end, one can solve for the terms $g_{ij}^{(4)}$ to the zeroth order of $a$ using only the general relativity (GR) part, i.e. using $R_{ij}^{(4)}$ and neglecting completely the contribution from ${\cal M}_{ij}^{(4)}$. At the end, we substitute the result in (\ref{system of expanded equations}) when trying to find  $g_{00}^{(4)}$.
\footnote{We would like to thank Clifford M. Will for bringing this point to our attention.}
This introduces an error of ${\cal O} \left(a^2\right)$ in our calculations. The solution of $g_{ij}^{(4)}$ in GR was first given by Chandrasekhar and Nutku \cite{chandrasekhar} in the PPN gauge.  However, to be consistent  we should get an answer that obeys the constraint  (\ref{constraint}).  We work out the details of these calculations in appendix E.  Using  the results of appendix E along with the dictionary of appendix F we find
\begin{eqnarray}
\nonumber
\stackrel{(4)}{R_{00}}&=&\nabla^2\left[-\frac{1}{2}\stackrel{(4)}{g_{00}}-\frac{1}{2}U^2+\frac{7/2+3a}{1+3a}\Phi_2+\frac{\Phi_W}{2(1+3a)}\right.\\
\nonumber
&&\left.-\frac{3a}{1+3a}\left({\cal A}+{\cal B}-\Phi_1 \right) \right]\\
\nonumber
\stackrel{(4)}{{\cal M}_{00}}&=&\nabla^2\left[\frac{1}{2}\stackrel{(4)}{g_{00}}-\frac{35}{16}U^2-\frac{11}{8}\Phi_{W}-\frac{3}{2}\Phi_1-\frac{23}{8}\Phi_2\right.\\
\nonumber
&&\left.+\frac{3}{2}{\cal A}+\frac{3}{2}{\cal B}+\frac{1}{16}U_{ij}U_{ij} \right]\\
\nonumber
\stackrel{(4)}{{\cal M}_{ii}}&=&\nabla^2\left[\frac{115}{16}U^2-\frac{1}{8}\Phi_W+6\Phi_1-\frac{69}{8}\Phi_2+8\Phi_3 \right.\\
&&\left. +2{\cal A}+2{\cal B}+\frac{7}{16}U_{ij}U_{ij}+{\cal V}\right]\,,
\end{eqnarray}
where ${\cal V}=E_{i,i}$, and $E_i$ are arbitrary  functions associated with $g_{ij}^{(4)}$ as  explained in appendix E.
Moreover, the r.h.s of the last equation of (\ref{system of expanded equations}) reads
\begin{eqnarray}
\nonumber
&&4\pi G \left(\stackrel{(4)}{T^{00}}+\stackrel{(4)}{T^{ii}}-2\stackrel{(2)}{g_{00}}\stackrel{(2)}{T^{00}} \right)=\left(1-\frac{9}{2}a\right)\times\\
&&\times\nabla^2\left[-2\Phi_1+2\Phi_2-\Phi_3-3\Phi_4 \right]\,.
\end{eqnarray}

Now using the last eq. in (\ref{system of expanded equations}), and solving for $g_{00}^{(4)}$ to the first order in $a$ we obtain
\begin{eqnarray}
\nonumber
\stackrel{(4)}{g_{00}}&=&\left(-1+\frac{15}{2}a \right)U^2+\left(1-\frac{5}{2}a \right)\Phi_W\\
\nonumber
&&+\left(4-\frac{11}{2}a \right)\Phi_1+\left(3-\frac{59}{2}a \right)\Phi_2\\
\nonumber
&&+2\Phi_3+\left(6-24a \right)\Phi_4-\frac{5}{2}a\left({\cal A}+{\cal B} \right)\\
\label{result for g004}
&&+\frac{a}{2}U_{ij}U_{ij}+a {\cal V}\,.
\end{eqnarray}

Next, we impose the constraint (\ref{cons condition}), and derive the equation that determines the function ${\cal V}$.

\subsection{Determining ${\cal V}$}

To find the condition that determines ${\cal V}$, we use the constraint  (\ref{cons condition}) to ${\cal O}(4)$, as in the last eq. of (\ref{expanded constraint}). Taking the derivative of the aforementioned equation with respect to $x^i$, and using eqs. (\ref{gij for GR}), (\ref{integrability condition}), and (\ref{chandra solution}) we obtain
\begin{eqnarray}
\nonumber
&&\nabla^4{\cal V}+\frac{1}{4}\nabla^2 S_{ii}-\stackrel{(3)}{{\cal M}_{0i,0i}}-\left(\stackrel{(2)}{\Gamma_{jj}^k}\stackrel{(2)}{{\cal M}_{ki}}+\stackrel{(2)}{\Gamma_{kj}^k}\stackrel{(2)}{{\cal M}_{ij}}\right.\\
\nonumber
&&\left.+\stackrel{(2)}{g_{jk}}\stackrel{(2)}{{\cal M}_{ij,k}}-\stackrel{(2)}{\Gamma_{00}^k} \stackrel{(2)}{{\cal M}_{ki}}-\stackrel{(2)}{\Gamma_{0i}^0} \stackrel{(2)}{{\cal M}_{00}} \right)_{,i}\\
\label{equation for V}
&&+\stackrel{(4)}{{\cal B}_{ij,ij}}+\stackrel{(4)}{{\cal P}_{ij,ij}}=0\,,
\end{eqnarray}
where $S_{ii}$ is given by (\ref{the value of Sii}), and  the rest of the quantities are given in appendix C. Using the dictionary in (\ref{nabla 4}), and solving for ${\cal V}$ we find after long, yet straightforward calculations
\begin{eqnarray}
\nonumber
{\cal V}&=&-\frac{83}{16}U^2-\frac{5}{4}\Phi_W-\frac{45}{8}\Phi_1+\frac{7}{2}\Phi_2-2\Phi_3+\frac{33}{4}\Phi_4\\
\label{the solution for V}
&&+\frac{5}{4}\Phi_5-\frac{1}{2}{\cal A}+\frac{29}{8}{\cal B}-\frac{1}{16}U_{ij}U_{ij}\,.
\end{eqnarray}

In the next section we read off the PPN parameters and constraint the value of $a$.

\section{PPN parameter values and interpretation}

To extract the PPN parameters one has to bring the the form of the metric to the standard PPN metric by means of a gauge transformation. However, since we are dealing with a theory that breaks the diffeomorphism invariance, one does not expect that such transformation will always respect the constraint (\ref{cons condition}). We can overcome this problem, simply, by transforming the standard PPN metric to the gauge that satisfies (\ref{cons condition}). This transformation is given by
\begin{eqnarray}
\nonumber
g_{ij}^{\mbox{\scriptsize PPN}}&=&g_{ij}-2\lambda_2\chi_{,ij}\\
\nonumber
g_{0i}^{\mbox{\scriptsize PPN}}&=&g_{0i}-(\lambda_1+\lambda_2)(V_i-W_i)\\
\nonumber
g_{00}^{\mbox{\scriptsize PPN}}&=&g_{00}-2\lambda_2\left(U^2+\Phi_W-\Phi_2  \right)-2\lambda_1\left ({\cal A}+{\cal B}-\Phi_1\right)\,,\\
\label{the PPN gauge transformation}
\end{eqnarray}
where the expressions for $g_{\mu\nu}^{\mbox{\scriptsize PPN}}$ are functions of ten PPN parameters as given in \cite{Will:1993ns}. Comparing eqs. (\ref{the PPN gauge transformation}) and (\ref{result for g004}) we can read off the values of the gauge parameters $\lambda_1$ and $\lambda_2$, as well as the PPN parameters as given in table \ref{PPN parameters}.

\begin{table}
\centerline{
\begin{tabular}{|c|c|c|c|}
        \hline
     parameter  & value &  effect     & limit               \\ \hline \hline
     $\gamma-1$ & -3a   &  time delay & $2.3\times 10^{-5}$  \\ \hline
                &       & light deflection & $4\times 10^{-4}$\\\hline
     $\beta-1$  &$-\frac{85}{32}a$ & perihelion shift & $3\times 10^{-3}$ \\\hline
                &                  & Nordtvedt effect  & $2.3\times 10^{-4}$ \\ \hline
     $\xi$      &$\frac{3}{8}a$      &  earth tides      & $10^{-3}$             \\ \hline
     $\alpha_1$ & $0$                   & orbital polarization& $10^{-4}$            \\\hline
     $\alpha_2$ & $0$                   & orbital polarization& $4\times10^{-7}$            \\\hline
     $\alpha_3$ & $-\frac{65}{8}a$       & orbital polarization& $4\times10^{-20}$            \\\hline
     $\zeta_1$ & $\frac{39}{8}a$       &--- & $2\times10^{-2}$            \\\hline
     $\zeta_2$ & $-\frac{179}{16}a$       &binary acceleration& $4\times10^{-5}$            \\\hline
      $\zeta_3$ & $-a$       &Newton's 3rd law& $10\times10^{-8}$            \\\hline
      $\zeta_4$ & $\frac{5}{8}a$ &--- & --- \\\hline
\end{tabular}
}
\caption{The values and limits on the PPN parameters \cite{will living review}.}
\label{PPN parameters}
\end{table}

The parameters $\gamma$ and $\beta$ are Eddington-Robertson-Schiff parameters used to describe the classical tests of theories of gravity; namely the deflection of light, time delay and  perihelion shift. The parameter $\xi$ is non-zero in any theory of gravity that predicts preferred-location effects. Also, $\alpha_1$, $\alpha_2$ and $\alpha_3$ measure whether or not
the theory predicts post-Newtonian preferred-frame effects.
\footnote{The reader can refer to \cite{will living review} for a recent and extended review of the different tests of GR.}

When one attempts to devise integral conservation laws, we search for a quantity $\Theta^{\mu\nu}$ which reduces to $T^{\mu\nu}$ in flat spacetime and satisfies $\partial_{\mu}\Theta^{\mu\nu}=0$. It was shown in \cite{Lee:1974nq} that such quantity can exist only if all the parameters $\{ \alpha_3,\zeta_1,\zeta_2,\zeta_3,\zeta_4\}$ are identically zero. Non-zero values of these parameters measure the extent at which a theory of gravity predicts violations of conservation of total energy and momentum. Notice that the parameter $\alpha_3$ plays a dual role, both as a conservation law and preferred-frame parameter.

A bound on $\alpha_3$, of  $4\times 10^{-20}$  was reported in \cite{Bell:1996ir,Stairs:2005hu} from the period derivatives of $21$ millisecond pulsars. This small bound puts severe constraint on the value of $\{a_i\}$, and in turn on the diffeomorphism-violating Lagrangians.

Finally, it is worth noting that our result for $g_{00}$ to ${\cal O}(4)$ contains additional potentials $U_{ij}U_{ij}$ and $\Phi_5$ that are not present in the standard PPN formalism. It would be interesting to devise experiments aimed to measure the effects of such terms in gravitational systems.

\section{Summary}

Our work helps to quantify the physical content of general covariance. We have allowed for the possibility of small violations of diffeomorphism invariance through a class of operators with two derivatives of the metric. An analysis to linear order produced modest constraints from light bending. However, we have used the PPN formalism to bound the non-invariance of a sample operator in this basis, which produces a far stronger constraint. By far the strongest result comes from the absence of preferred frame effects in pulsars, and leads to the constraint that the dimensionless parameter $a$ must be less than $10^{-20}$ of gravitational strength.

Tests of diffeomorphism invariance are of interest in its own right. We want to have quantitative probes of this fundamental property. Moreover, we have argued that this constraint is relevant for theories in which general relativity is an emergent phenomenon from a more fundamental theory that lacks a fundamental version of diffeomorphism invariance. There are many directions that extensions of this initial investigation can go and we feel that the topic deserves further study.

\section*{Acknowledgements}

This work has been supported in part by the NSF grants PHY- 055304 and PHY - 0855119, and in part by the Foundational Questions
Institute. We would like to thank T.J. Blackburn, Luca Grisa and Lorenzo Sorbo for interesting discussions. Also, we would like to thank Clifford Will for useful communications.

\appendix
\renewcommand{\theequation}{A\arabic{equation}}
  \setcounter{equation}{0}  
  \section*{Appendix A: Operators of the linearized theory}

In this appendix we write down the lower, marginal and higher dimensional operators for a linearized theory of gravity where we expand $g_{\mu\nu}=\eta_{\mu\nu}+h_{\mu\nu}$ .

\begin{itemize}

\item Dimension-2 operators\\
The lowest dimension operators can only contain powers of the field, without derivatives. At leading order,
\begin{equation}
h^2\,,h^{\mu\nu}h_{\mu\nu}
\end{equation}
The Pauli-Fierz mass term is
\begin{equation}
\frac{m^2}{4}(h^{\mu\nu}h_{\mu\nu}-h^2)
\end{equation}

\item Dimension-3 operators

\begin{eqnarray}
h^3\,,h h^{\alpha}_{\beta}h_{\alpha}^{\beta}\,,h^{\mu}_{\alpha}h^{\alpha}_{\nu}h^{\nu}_{\mu}
\end{eqnarray}

\item Dimension-4 operators\\
At dimension-4 we start to have the possibility of operators with two derivatives, starting the next series in the derivative expansion. Those without derivatives are

\begin{eqnarray}
h^4\,,h^2 h^{\alpha}_{\beta}h^{\beta}_{\alpha}\,, h h^{\alpha}_{\beta}h^{\beta}_{\gamma}h^{\gamma}_{\alpha}\,,h^{\alpha}_{\beta}h^{\beta}_{\gamma}h^{\gamma}_{\delta}h^{\delta}_{\alpha}
\end{eqnarray}
while those with two derivatives are the set
\begin{eqnarray}
\nonumber
{\cal C}^{(4)}&=&\{\partial_{\mu}h^{\alpha\beta}\partial^{\mu}h_{\alpha\beta}\,,\partial_{\alpha}h^{\alpha\beta}\partial_{\gamma}h^{\gamma}_{\beta}\,,\partial_{\alpha}h \partial^{\alpha}h \,,\partial_{\alpha}h\partial^{\beta}h^{\alpha}_{\beta} \}\\
\end{eqnarray}

\item Dimension-5 operators\\
At this order we stop listing the series with zero derivatives and show the next order terms in the series with two derivatives. These all have three powers of the field and occur in the combinations
\begin{equation}
{\cal C}^4 h\,,~~{\cal C}^{(4)\alpha\beta}h_{\alpha\beta}
\end{equation}
where ${\cal C}^4$ is defined above and
\begin{eqnarray}
\nonumber
{\cal C}^{(4)\alpha\beta}=\{&&\partial^{\alpha}h^{\mu\nu}\partial^{\beta}h_{\mu\nu}\,,\partial^{\alpha}h\partial^{\beta}h\,,\partial_{\mu}h^{\alpha\beta}\partial^{\mu}h\,,\\
\nonumber
&&\partial^{\nu}h^{\alpha\beta}\partial^{\mu}h_{\mu\nu}\,,\partial^{\nu}h^{\alpha}_{\mu}\partial_{\nu}h^{\mu\beta}\,,\partial_{\nu}h^{\alpha\nu}\partial_{\mu}h^{\mu\beta}\,,\\
\nonumber
&&
\partial^{\alpha}h^{\mu\nu}\partial_{\nu}h^{\beta}_{\mu}\,,\partial^{\alpha}h\partial^{\mu}h^{\beta}_{\mu}\,,\partial^{\mu}h\partial^{\alpha}h^{\beta}_{\mu}\,,\\
&&\partial_{\nu}h^{\mu\nu}\partial^{\alpha}h^{\beta}_{\mu} \}
\end{eqnarray}

\item Dimension-6 operators\\
At sixth order, the four derivative series starts. We here list only those with four derivatives and two powers of the field.
\begin{eqnarray}
\nonumber
&&\partial_{\alpha}\partial_{\beta}h\partial^{\alpha}\partial^{\beta}h\,,\partial_{\alpha}\partial_{\beta}h_{\mu\nu}\partial^{\alpha}\partial^{\beta}h^{\mu\nu}\,,\partial_{\alpha}\partial_{\beta}h^{\mu\nu}\partial_{\mu}\partial_{\nu}h^{\alpha\beta}\,,\\
&&\partial_{\alpha}\partial_{\beta}h_{\mu\nu}\partial^{\alpha}\partial^{\nu}h_{\beta}^{\mu}\,.
\end{eqnarray}

\end{itemize}

\appendix
\renewcommand{\theequation}{B\arabic{equation}}
  \setcounter{equation}{0}  
  \section*{Appendix B: Gravitational potentials}

In this appendix, we list the various gravitational potentials used to construct the metric.
\begin{eqnarray}
\nonumber
U&=&\int d^3x' \frac{\rho'}{|\vec x-\vec x'|}\\
\nonumber
U_{ij}&=&\int d^3 x' \frac{\rho'\left(x-x' \right)_i\left(x-x'\right)_j}{|\vec x-\vec x'|^3}\\
\nonumber
V_i&=&\int d^3 x' \frac{\rho' v_i'}{|\vec x-\vec x'|}\\
\nonumber
W_{i}&=&\int d^3 x' \frac{\rho' \vec v'\cdot\left(\vec x-\vec x' \right)\left(x-x' \right)_i}{|\vec x-\vec x'|^3}\\
\nonumber
\Phi_1&=&\int d^3x' \frac{\rho' v'^2}{|\vec x-\vec x'|}\,,\Phi_2=\int d^3x' \frac{\rho' U'}{|\vec x-\vec x'|}\\
\nonumber
\Phi_3&=&\int d^3x' \frac{\rho' \Pi'}{|\vec x-\vec x'|}\,,\Phi_4=\int d^3x' \frac{p'}{|\vec x-\vec x'|}\\
\nonumber
{\cal A}&=& \int d^3x' \frac{\rho' \left[\vec v'\cdot\left(\vec x-\vec x'\right) \right]^2}{|\vec x-\vec x'|^3}\\
\nonumber
{\cal B}&=& \int d^3x' \frac{\rho' d\vec v'/dt\cdot\left(\vec x-\vec x'\right)}{|\vec x-\vec x'|}\\
\nonumber
\Phi_W&=&\int d^3 x'\rho'\rho''\frac{\vec x-\vec x'}{|\vec x-\vec x'|^3}\cdot\left(\frac{\vec x'-\vec x''}{|\vec x-\vec x''|}-\frac{\vec x-\vec x''}{|\vec x'-\vec x''|} \right)\,.\\
\end{eqnarray}
These potentials satisfy the differential relations
\begin{eqnarray}
\nonumber
&&\nabla^2 V_i=-4\pi\rho v_i\,,\quad V_{i,i}=-U_{,0}\\
\nonumber
&&\nabla^2\Phi_1=-4\pi\rho v^2\,,\quad \nabla^2\Phi_2=-4\pi \rho U\\
\nonumber
&&\nabla^2\Phi_3=-4\pi \rho \Pi\,,\quad \nabla^2\Phi_4=-4\pi p\\
\nonumber
&&\nabla^2\left(\Phi_W+2U^2-3\Phi_2 \right)=2\chi_{,ij}U_{,ij}\\
&&\chi_{,00}={\cal A}+{\cal B}-\Phi_1\,,
\end{eqnarray}
where
\begin{eqnarray}
\nonumber
\chi&=&-\int d^3 x' \rho'|\vec x-\vec x' |\\
\chi_{,ij}&=&-\delta_{ij}U+U_{ij}\,,\quad \nabla^2\chi=-2U\,.
\end{eqnarray}

In addition, consider the potential
\begin{equation}
\psi_{i}=\int d^3 x'\frac{U'_{ij}\rho'_{,j}}{|\vec x-\vec x'|}\,,
\end{equation}
such that $\nabla^2 \psi_{i}=-4\pi U_{ij}\rho_{,j}$. Hence, we define the potential $\Phi_5$ as
\begin{equation}
\nabla^2\psi_{i,i}\equiv\nabla^4\Phi_5=4\pi \rho_{,i}U_{,i}-4\pi U_{ij}\rho_{,ij}\,.
\end{equation}
%

\appendix
\renewcommand{\theequation}{C\arabic{equation}}
  \setcounter{equation}{0}  
  \section*{Appendix C: Expressions Used Throughout the Paper }

In this appendix we give the form of the different expressions used throughout this paper.

The components of the affine connections are
\begin{eqnarray}
\nonumber
\stackrel{(2)}{\Gamma^{i}_{00}}&=&-\frac{1}{2}\stackrel{(2)}{g_{00,i}}\,,\quad \stackrel{(4)}{\Gamma^{i}_{00}}=-\frac{1}{2}\stackrel{(4)}{g_{00,i}}+\stackrel{(3)}{g_{0i,0}}+\frac{1}{2}\stackrel{(2)}{g_{ij}}\stackrel{(2)}{g_{00,j}}\\
\nonumber
\stackrel{(3)}{\Gamma^{i}_{0j}}&=&\frac{1}{2}\left[\stackrel{(3)}{g_{i0,j}}+\stackrel{(2)}{g_{ij,0}}-\stackrel{(3)}{g_{j0,i}} \right]\,,\quad \stackrel{(3)}{\Gamma_{00}^{0}}=-\frac{1}{2}g_{00,0}^{(2)} \\
\nonumber
\stackrel{(2)}{\Gamma^{i}_{jk}}&=&\frac{1}{2}\left[\stackrel{(2)}{g_{ij,k}}+\stackrel{(2)}{g_{ik,j}}-\stackrel{(2)}{g_{jk,i}} \right]\,,\quad \stackrel{(2)}{\Gamma_{0i}^{0}}=-\frac{1}{2}\stackrel{(2)}{g_{00,i}}\\
\nonumber
\stackrel{(4)}{\Gamma^{i}_{jk}}&=&\frac{1}{2}\delta^{ip}\left[\stackrel{(4)}{g_{pj,k}}+\stackrel{(4)}{g_{pk,i}}-\stackrel{(4)}{g_{ik,p}} \right]\\
\label{affine connections}
&&-\frac{1}{2}\stackrel{(2)}{g_{ip}}\left[\stackrel{(2)}{g_{pi,k}}+\stackrel{(2)}{g_{pk,i}}-\stackrel{(2)}{g_{ik,p}} \right]\,.
\end{eqnarray}

The lower order expressions for $R$ and ${\cal M}$ read
\begin{eqnarray}
\nonumber
 \stackrel{(2)}{R_{00}}&=&-\frac{1}{2}\stackrel{(2)}{g_{00,ii}}\,,\quad\stackrel{(2)}{{\cal M}_{00}}=\frac{3}{2} \stackrel{(2)}{g_{00,ii}}\\
 \nonumber
  \stackrel{(2)}{R_{ij}}&=&\frac{1}{2}\left[\stackrel{(2)}{g_{00,ij}} -\stackrel{(2)}{g_{kk,ij}}+\stackrel{(2)}{g_{ik,kj}}+\stackrel{(2)}{g_{kj,ki}}-\stackrel{(2)}{g_{ij,kk}} \right]\\
  \nonumber
  \stackrel{(2)}{{\cal M}_{ij}}&=&\frac{1}{2}\left[-\stackrel{(2)}{g_{ik,kj}}-\stackrel{(2)}{g_{jk,ki}} +3 \stackrel{(2)}{g_{ij,kk}}      \right]\\
  \nonumber \stackrel{(3)}{R_{i0}}&=&\frac{1}{2}\left[-\stackrel{(2)}{g_{jj,0i}}+\stackrel{(3)}{g_{j0,ij}}+\stackrel{(2)}{g_{ij,j0}}-\stackrel{(3)}{g_{i0,kk}}\right]\\
\stackrel{(3)}{{\cal M}_{i0}}&=&\frac{3}{2}\stackrel{(2)}{g_{i0,kk}}\,.
\label{linearized set}
\end{eqnarray}

The higher order perturbations of $R$ and ${\cal M}$  are given by (the constraint (\ref{constraint}) being imposed)
\begin{eqnarray}
\nonumber
\stackrel{(4)}{R_{00}}&=&-\frac{1}{2}\stackrel{(2)}{g_{ii,00}}+\stackrel{(3)}{g_{i0,i0}}-\frac{1}{2}\stackrel{(4)}{g_{00,kk}}+\frac{1}{2}\stackrel{(2)}{g_{ij}}\stackrel{(2)}{g_{00,ij}}\\
\nonumber
&&-\frac{1}{4}\stackrel{(2)}{g_{00,i}}\stackrel{(2)}{g_{00,i}}-\frac{1}{4}\stackrel{(2)}{g_{00,i}}\stackrel{(2)}{g_{jj,i}}\\
\nonumber
\stackrel{(4)}{{\cal M}_{00}}&=&-\frac{15}{2}\stackrel{(2)}{g_{00,i}}\stackrel{(2)}{g_{00,i}}+3\stackrel{(2)}{g_{ii,j}}\stackrel{(2)}{g_{00,j}}-2\stackrel{(2)}{g_{00,00}}+2\stackrel{(4)}{g_{00,jj}}\\
\nonumber
&&-4\stackrel{(3)}{g_{0j,0j}}-6\stackrel{(2)}{g_{ij}}\stackrel{(2)}{g_{00,ij}}+\frac{3}{2}\stackrel{(2)}{g_{ij,k}}\stackrel{(2)}{g_{ij,k}}-\stackrel{(2)}{g_{jk,i}}\stackrel{(2)}{g_{ij,k}}\\
\nonumber
\stackrel{(4)}{{\cal M}_{ii}}&=&-\frac{3}{4}\stackrel{(2)}{g_{ii,j}}\stackrel{(2)}{g_{00,j}}+\frac{3}{4}\stackrel{(2)}{g_{kk,j}}\stackrel{(2)}{g_{ii,j}}-\frac{3}{2}\stackrel{(2)}{g_{ii,00}}+\stackrel{(3)}{g_{0i,0i}}\\
\nonumber
&&-\frac{3}{8}\stackrel{(2)}{g_{00,i}}\stackrel{(2)}{g_{00,i}}+\frac{9}{8}\stackrel{(2)}{g_{ik,j}}\stackrel{(2)}{g_{ik,j}}-\frac{3}{4}\stackrel{(2)}{g_{ij,k}}\stackrel{(2)}{g_{kj,i}}\\
\label{higher order terms}
&&-\frac{3}{2}\stackrel{(2)}{g_{kp}}\stackrel{(2)}{g_{ii,kp}}+\frac{1}{2}\left( 3\stackrel{(4)}{g_{ii,kk}}-2\stackrel{(4)}{g_{ik,ik}}\right)\,,
\end{eqnarray}
where the summation is implied in $\stackrel{(4)}{{\cal M}_{ii}}$.

The functions ${\cal B}_{ij}$ and ${\cal P}_{ij}$ defined in (\ref{equation for V}) are given by
\begin{eqnarray}
\nonumber
\stackrel{(4)}{{\cal B}_{ij}}&=&-\delta_{ij}\left(\frac{3}{8}\stackrel{(2)}{g_{00,k}}\stackrel{(2)}{g_{00,k}}+\frac{3}{8}\stackrel{(2)}{g_{nk,m}}\stackrel{(2)}{g_{nk,m}}\right.\\
\nonumber
&&\left.-\frac{1}{4}\stackrel{(2)}{g_{nm,k}}\stackrel{(2)}{g_{km,n}}  \right)+ \stackrel{(2)}{\Gamma_{00}^i}\stackrel{(2)}{\Gamma_{00}^j}+\stackrel{(2)}{\Gamma_{km}^i}\stackrel{(2)}{\Gamma_{km}^j}\\
\nonumber
&&+2\stackrel{(2)}{\Gamma_{ik}^m}\stackrel{(2)}{\Gamma_{jk}^m}+2\stackrel{(2)}{\Gamma_{i0}^0}\stackrel{(2)}{\Gamma_{j0}^0}\\
\nonumber
\stackrel{(2)} {{\cal P}_{ij}}&=&\stackrel{(2)}{\Gamma_{0k}^0}\stackrel{(2)}{A_{ij}^k}+\stackrel{(2)}{\Gamma_{km}^m}\stackrel{(2)}{A_{ij}^k}+\stackrel{(3)}{A_{ij,0}^0}+\stackrel{(4)}{{\cal Q}^k_{ij,k}}\\
\nonumber
\stackrel{(4)}{{\cal Q}^k_{ij}}&=& -\frac{1}{2}\stackrel{(2)}{g_{ip}}\left(\stackrel{(2)}{g_{pj,k}}+\stackrel{(2)}{g_{pk,j}}-\stackrel{(2)}{g_{jk,p}}   \right)\\
\nonumber
&&-\frac{1}{2}\stackrel{(2)}{g_{jp}}\left(\stackrel{(2)}{g_{pi,k}}+\stackrel{(2)}{g_{pk,i}}-\stackrel{(2)}{g_{ik,p}}   \right)\\
\nonumber
&&+\frac{1}{2}\stackrel{(2)}{g_{kp}}\left(\stackrel{(2)}{g_{pi,j}}+\stackrel{(2)}{g_{pj,i}}-\stackrel{(2)}{g_{ij,p}}   \right)\\
\nonumber
&&+\left( \stackrel{(2)}{g_{ni}}\stackrel{(2)}{\Gamma_{jk}^n}+\stackrel{(2)}{g_{nj}}\stackrel{(2)}{\Gamma_{ik}^n}-\stackrel{(2)}{g_{km}}\stackrel{(2)}{\Gamma_{jm}^i}-\stackrel{(2)}{g_{km}}\stackrel{(2)}{\Gamma_{im}^j}  \right)\,.\\
\label{definition of P AND B}
\end{eqnarray}
%

\appendix
\renewcommand{\theequation}{D\arabic{equation}}
  \setcounter{equation}{0}  
  \section*{Appendix D: Linearized Version of the Equations of Motion}

One can use the linearized version of the equations of motion (\ref{com linearized equation of motion}) along with the constraint $\partial_{\beta}h^{\alpha\beta}=0$ to solve for $g_{ij}$ to ${\cal O}(2)$ and $g_{0i}$ to ${\cal O}(3)$.
\footnote{Notice that $h_{00}= g_{00}^{(2)}$, $h_{ij}=g_{ij}^{(2)}$, and $h_{0i}=g_{0i}^{(3)}$.}
To this end, we take the trace of (\ref{com linearized equation of motion}), and write $h$ in terms of $T$ to find
\begin{eqnarray}
\nonumber
&&\left(1+a_1-3a_3 \right)\square h^{\alpha\beta}+\left(1-a_2+2a_4  \right)\partial^{\alpha}\partial^{\beta}h\\
\label{simple form of linear eq}
&&=-16\pi G\left(\stackrel{(2)}{T^{\alpha\beta}}-A\eta^{\alpha\beta}\stackrel{(2)}{T}  \right)
\end{eqnarray}
where
\begin{equation}
A=\frac{1-a_2+a_4+a_5}{2-a_1-3a_2+3a_3+2a_4+4a_5}\,.
\end{equation}

Solving for the $h_{00}$ component we get
\begin{equation}
h_{00}=2\alpha U\,,
\end{equation}
where
\begin{equation}
\alpha=\frac{2G\left(1-a_1-2a_2+3a_3+a_4+3a_5\right)}{\left(1+a_1-3a_3\right)\left(2-a_1-3a_2+3a_3+2a_4+4a_5\right)}\,.
\end{equation}

The most general solution of $h_{ij}$ is given by
\begin{equation}
h_{ij}=\sigma_1 \delta_{ij} U+\sigma_2U_{ij}\,,
\end{equation}
where $\sigma_1$ and $\sigma_2$ are constants to be determined. Using the constraint $\partial_{i}h_{ij}$ we find $\sigma_1=\sigma_2$. Then, substituting into eq. (\ref{simple form of linear eq}) we find that the following two equations
\begin{equation}
2\sigma_1(1+a_1-3a_3)=\frac{4G\left(1-a_2+a_4+a_5 \right)}{2-a_1-3a_2+3a_3+2a_4+4a_5}\,,
\end{equation}
and
\begin{equation}
\sigma_1\left(-1+a_1-3a_3+2a_2-4a_4  \right)=-\alpha\left(1-a_2+2a_4 \right)\,,
\end{equation}
have to be satisfied simultaneously. This can be true only if we take $a_4=a_5$. This is exactly what we found before in eq. (\ref{linearized constraint}). Setting $\alpha=1$, we obtain the normalized value of $\sigma_1$
\begin{equation}
\sigma_1=\frac{1-a_2+2a_4}{1-a_1-2a_2+3a_3+4a_4}\,,
\end{equation}
from which we immediately read the PPN parameter $\gamma$
\begin{equation}
\gamma=\frac{1-a_2+2a_4}{1-a_1-2a_2+3a_3+4a_4}\,,
\end{equation}
which reduces in the limiting case $a_1=a_2...=a_5=0$ to the GR result. The deflection of light is proportional to $\gamma+1$, which gives (\ref{propagator light bending}).

\appendix
\renewcommand{\theequation}{E\arabic{equation}}
  \setcounter{equation}{0}  
  \section*{Appendix E:  Solving  for $\left( 3\stackrel{(4)}{g_{ii,kk}}-2\stackrel{(4)}{g_{ik,ik}}\right)$}

To find $g_{ij}$ to ${\cal O}(4)$, we need to write the $i-j$ component of the equation of motion (\ref{conv form}) to ${\cal O}(4)$. However, since we are interested in $g_{00}^{(4)}$ to first order in $a$, we can solve for $g_{ij}^{(4)}$ to zeroth order in $a$. Hence, the  $i-j$ component of  (\ref{conv form}) reduces to the GR result $R_{ij}+{\cal O}(a)=8\pi G \left(T_{ij}-g_{ij}T/2 \right)$. Moreover, the solution of $g_{ij}^{(4)}$ should respect the constraint (\ref{constraint}).

The $i-j$ component of the Ricci tensor can be evaluated to the fourth order to find
\begin{eqnarray}\label{gij for GR} \stackrel{(4)}{g_{ij,kk}}+\stackrel{(4)}{g_{kk,ij}}-\stackrel{(4)}{g_{ik,kj}}-\stackrel{(4)}{g_{jk,ki}}=S_{ij}\,
\end{eqnarray}
where $S_{ij}$ is a complicated function of the various potentials. Contracting eq. (\ref{gij for GR}) results in
\begin{equation}\label{equation 1}
\nabla^2 \stackrel{(4)}{g_{ii}}-\stackrel{(4)}{g_{ij,ij}}=\frac{1}{2}S_{ii}\,
\end{equation}
while differentiating it with respect to $j$ gives
\begin{equation}\label{equation 2}
\left(\nabla^2 \stackrel{(4)}{g_{ii}}-\stackrel{(4)}{g_{ij,ij}}\right)_{,m}=S_{mk,k}\,.
\end{equation}
From eqs. (\ref{equation 1}) and (\ref{equation 2}) we obtain the integrability condition
\begin{equation}\label{integrability condition}
\left(S_{ij}-\frac{1}{2}\delta_{ij}S_{kk}\right)_{,i}=0\,.
\end{equation}
It was shown by Chandrasekhar and Nutku \cite{chandrasekhar} that this condition is indeed satisfied in GR in the PPN gauge. Since GR is a diffeomorphism-invariant theory, we conclude that this condition still holds when using the constraint (\ref{constraint}). It was also shown in [] that the solution of eq.(\ref{equation 1}) is given by
\begin{equation}\label{chandra solution}
\nabla^2\stackrel{(4)}{g_{ij}}=S_{ij}+E_{i,j}+E_{j,i}\,,
\end{equation}
where $E_i$ are arbitrary functions.
\footnote{Strictly speaking, since we are using only GR, the constraint (\ref{constraint}) is not more than a gauge fixing to ${\cal O}(2)$, and $E_i$ accounts for the gauge freedom to ${\cal O}(4)$ . Although we are free to fix a gauge and hence the values of  $E_i$ in case we were dealing only with GR, the values of these functions will be determined upon using the consistency condition (\ref{cons condition}) to ${\cal O}(4)$. }
Now, using eqs. (\ref{equation 1}) and  (\ref{chandra solution}) we get
\begin{equation}
 3\stackrel{(4)}{g_{ii,kk}}-2\stackrel{(4)}{g_{ik,ik}}=2S_{ii}+2E_{i,i}\,
\end{equation}
where
%
\begin{eqnarray}
\nonumber
S_{ii}&=&\frac{1}{2}\stackrel{(2)}{g_{00,i}}\stackrel{(2)}{g_{00,i}}+\stackrel{(2)}{g_{00}}\stackrel{(2)}{g_{00,ii}}+\frac{3}{2}\stackrel{(2)}{g_{kp,i}}\stackrel{(2)}{g_{kp,i}}+\stackrel{(2)}{g_{kp}}\stackrel{(2)}{g_{kp,ii}}\\
\nonumber
&&+2\stackrel{(2)}{g_{ii,00}}+\stackrel{(2)}{g_{kp}}\stackrel{(2)}{g_{ii,pk}}-\stackrel{(2)}{g_{ik,m}}\stackrel{(2)}{g_{im,k}}-\frac{1}{2}\stackrel{(2)}{g_{mm,k}}\stackrel{(2)}{g_{ii,k}}\\
\nonumber
&&+\frac{1}{2}\stackrel{(2)}{g_{00,k}}\stackrel{(2)}{g_{ii,k}}-2\stackrel{(3)}{g_{0i,0i}}+\stackrel{(4)}{g_{00,kk}}\\
\label{the value of Sii}
&&-8\pi\left (-\stackrel{(4)}{T^{ii}}+3\stackrel{(4)}{T^{00}}-3\stackrel{(2)}{g
_{00}}\stackrel{(2)}{T^{00}}+\stackrel{(2)}{g
_{ii}}\stackrel{(2)}{T^{00}}  \right)\,.
\end{eqnarray}
%

The value of $\stackrel{(4)}{g_{00}}$ can be calculated using $R_{00}+{\cal O}(a)=8\pi G \left(T_{00}-g_{00}T/2 \right)$, and imposing the constraint (\ref{constraint}) to find
\begin{equation}
\stackrel{(4)}{g_{00}}=-U^2+\Phi_W+4\Phi_1-4\Phi_2+2\Phi_3+6\Phi_4\,.
\end{equation}
Hence, using the dictionary in appendix F we finally obtain
\begin{eqnarray}
\nonumber
S_{ii}=\nabla^2 \left[\frac{25}{4}U^2+\frac{5}{2}\Phi_W+8\Phi_1+\frac{15}{2}\Phi_2+8\Phi_3\right.\\
\left.+\frac{1}{4}U_{ij}U_{ij} \right]\,.
\end{eqnarray}
Writing $E_i$ as the sum of the gradient and  curl of a scalar and vector, i.e. $E_i={\cal V}_{,i}+(\nabla \times \vec A)_{,i}$, we find
\begin{equation}\label{der of E}
E_{i,i}=\nabla^2 {\cal V}\,.
\end{equation}

\appendix
\renewcommand{\theequation}{F\arabic{equation}}
  \setcounter{equation}{0}  
  \section*{Appendix F: Dictionary}

In this appendix, we give a dictionary for the different combinations that appear in our formalism.

Using the differential relations in appendix B, we obtain to the zeroth order of $a$
\begin{eqnarray}
\nonumber
\stackrel{(2)}{g_{ij,k}}\stackrel{(2)}{g_{ij,k}}&=&\nabla^2\left[\frac{9}{2}U^2+\Phi_W-7\Phi_2+\frac{1}{2}U_{ij}U_{ij} \right]\\
\nonumber
\stackrel{(2)}{g_{ij}}\stackrel{(2)}{g_{ij,kk}}&=&\nabla^2\left[-2U^2-\Phi_W+7\Phi_2 \right]\\
\nonumber
\stackrel{(2)}{g_{jk}}\stackrel{(2)}{g_{ii,jk}}&=&\nabla^2\left[4U^2+2\Phi_W+2\Phi_2\right]\\
\nonumber
\stackrel{(3)}{g_{0i,0i}}&=&\nabla^2\left[\Phi_1-{\cal A}-{\cal B} \right]\\
\nonumber
\stackrel{(2)}{g_{00,i}}\stackrel{(2)}{g_{00,i}}&=&\nabla^2\left[2U^2-4\Phi_2 \right]\\
\nonumber
\stackrel{(2)}{g_{ii,k}}\stackrel{(2)}{g_{jj,k}}&=&\nabla^2\left[8U^2-16\Phi_2 \right]\\
\nonumber
\stackrel{(2)}{g_{00,00}}&=&\nabla^2\left[\Phi_1-{\cal A} -{\cal B}  \right]\\
\nonumber
\stackrel{(2)}{g_{ik,j}}\stackrel{(2)}{g_{ij,k}}&=&\nabla^2\left[\frac{1}{2}U^2+\Phi_W+\Phi_2+\frac{1}{2}U_{ij}U_{ij}  \right]\\
\nonumber
\stackrel{(2)}{g_{00,i}}\stackrel{(2)}{g_{kk,i}}&=&\nabla^2\left[4U^2-8\Phi_2 \right]\\
\nonumber
\stackrel{(2)}{g_{00}}\stackrel{(2)}{g_{00,kk}}&=&\nabla^2\left[4\Phi_2\right]\\
\stackrel{(2)}{g_{ij}}\stackrel{(2)}{g_{00,ij}}&=&\nabla^2\left[2U^2+\Phi_W+\Phi_2\right]\,.
\end{eqnarray}
One can also show
\begin{eqnarray}
\nonumber
U_{,ij}U_{,ij}&=&\nabla^4\left(\frac{1}{4}U^2-\frac{1}{2}\Phi_2\right)+4\pi\rho_{,i}U_{,i}\\
\nonumber
\chi_{,ijk}U_{,ijk}&=&\nabla^4\left(\frac{3}{4}U^2+\frac{1}{4}\Phi
_W-\frac{5}{4}\Phi_2  \right)\\
\nonumber
&&+4\pi \rho_{,i}U_{,i}+2\pi U_{ij}\rho_{,ij}-2\pi U\nabla^2\rho\\
\nonumber
\chi_{,ijkm}\chi_{,ijkm}&=&\nabla^4\left(\frac{11}{4}U^2+\Phi_W-4\Phi_2+\frac{1}{4}U_{ij}U_{ij} \right)\\
\nonumber
&&+8\pi \rho_{,i}U_{,i}+4\pi U_{ij}\rho_{,ij}-4\pi U\nabla^2\rho\\
\nonumber
\nabla^4\Phi_2&=&-8\pi \rho_{,i}U_{,i}+16\pi^2\rho^2-4\pi U\nabla^2\rho\\
\nonumber
\nabla^4\Phi_4&=&\left(\frac{1}{2}\Phi_1-\frac{1}{2}\beta \right)
-4\pi\rho_{,i}U_{,i}+16\pi^2\rho^2\,.\\
\label{nabla 4}
\end{eqnarray}

\end{document}